\documentclass[12pt]{article} \usepackage{amsmath,amssymb}
\usepackage[dvips]{graphicx} \usepackage{feynmf}
\title{Physical Principles and Properties\\ of Unstable
  States
\thanks{Talk given at the CFIF Workshop on Time Asymmetric
    Quantum Theory, July 23-26, 2003, Lisbon, Portugal.}
}
\author{Piotr Kielanowski\\
  \textsl{\small Centro de Investigaci\'on y de Estudios Avanzados
    del IPN}\\
  \textsl{\small Ap. Postal 14-740, 0700 M\'exico D.F., Mexico}}
\date{} 
\begin{document}
\maketitle
\begin{abstract}
  The main subject of the paper is the description of unstable states
  in quantum mechanics and quantum field theory. Unstable states in
  quantum field theory can only be introduced as the intermediate
  states and not as asymptotic states. The absence of the intermediate
  unstable states from the asymptotic states is compatible with
  unitarity. Thus the concept of an unstable state is not introduced
  in quantum field theory despite the fact that an unstable state has
  well defined linear momentum, angular momentum and other intrinsic
  quantum numbers. In the rigged Hilbert space quantum mechanics one
  can define vectors that correspond to the unstable states. These
  vectors are the generalized eigenvectors (kets in the rigged Hilbert
  space) with complex eigenvalues of the self-adjoint Hamiltonian. The
  real part of the eigenvalue corresponds to
  the mass of an unstable state and the imaginary part is one half of
  the total width. Such vectors form the minimally complex semigroup
  representation of the Poincar\'e transformations into the forward
  light cone.
\end{abstract}
\section{Introduction}
Conventional quantum mechanics can be applied to the physical systems
that consist of stable particles. For such systems the space of the
states is the Hilbert space and the observables are represented by the
self adjoint operators in the Hilbert space.  This mathematical
representation is supplemented by the probabilistic interpretation of
quantum mechanics.

The necessity of including also the unstable states in quantum
mechanics appeared from the very beginning in the relation with
the phenomenon of the radioactivity and in particular $\alpha$
decay. The description of $\alpha$ decay was first formulated in
1928 by George Gamow~\cite{ref1} and was based on the calculation
of the transition probability of a potential barrier and led to
the appearance of the eigenfunctions of the Hamiltonian with
complex eigenvalues.  This problem can also be formulated in the
language of the quasi-stationary states in scattering
theory~\cite{ref2} where the $S$-matrix has simple poles on the
second sheet of the complex energy~\cite{ref3,ref4}. The most
important feature of such a formulation is that the partial cross
section in the vicinity of the pole is universal and mostly
independent of the details of the interaction potential and
depending predominantly on the position of the $S$-matrix pole.
This cross section is given by the Breit-Wigner formula which
depends on two parameters: $E_{0}$ and $\Gamma$ and a slowly
varying background function. $E_{0}$ is the energy of the
resonance state and $\Gamma$ is the width of the resonance state.

The vectors that correspond to the resonance states do not belong to
the Hilbert space. In the case of the complex energy eigenstate the
time dependent position wave function diverges for
$r\rightarrow\infty$ and cannot be normalized~\cite{ref5,ref4}. The
$S$-matrix in quantum mechanics is used for the description of
scattering. If in a given physical system there are bound states then
the $S$-matrix has simple poles for imaginary momenta $\Im(k)>0$ and
the bound state energy is equal to
\begin{equation}\label{eq1}
  E=\frac{\hbar^{2}k^{2}}{2m}.
\end{equation}
Unstable states correspond to poles of the $S$-matrix but they are
situated in the lower semi plane of complex momentum in the
vicinity of the imaginary axis. For the complex energy the
$S$-matrix bound state poles are on the first, physical sheet and
the resonance poles are on the second sheet below and above the
cut along the positive real axis~\cite{ref3,ref4}.

The eigenvalues of the Hamiltonian on the negative real axis
correspond to the bound states. The complex eigenvalues on the
second sheet correspond to the unstable states. In the Hilbert
space the Hamiltonian is self adjoint and cannot have complex
eigenvalues. It thus follows that the description of the unstable
states by eigenvectors in Hilbert space is not possible. Thus the
inclusion of the unstable states calls for a modification of
quantum theory. In my talk I will consider three kinds of
description of unstable states
\begin{enumerate}
\item Unstable states in quantum mechanics--Gamow states.
\item Unstable states in quantum field theory.
\item Rigged Hilbert space of unstable states.
\end{enumerate}
I will present the main assumptions of each method and their main
conclusions and will show the relation between the first two
approaches and the rigged Hilbert space description.
\section{Unstable states in quantum mechanics--Ga\-mow states}
\begin{figure}
  \centering \includegraphics[width=0.8\textwidth]{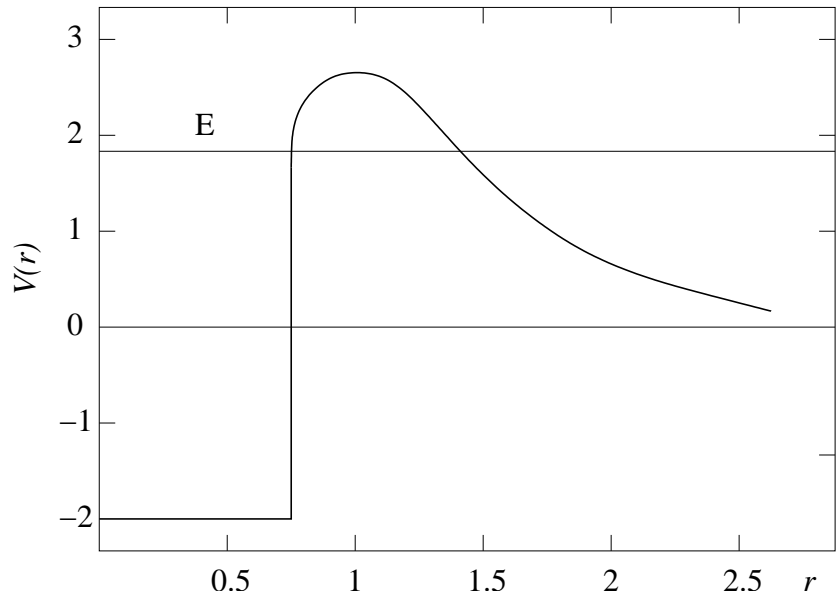}
  \caption{Potential $V(r)$ for a case with bound states and
    resonances.}
  \label{fig1}
\end{figure}

The quasi-stationary states in quantum mechanic appears when one
considers, e.g., the potentials that have finite width. If one
considers the three dimensional potential well shown in
Fig.~\ref{fig1} then for energies inside the potential well there may
exist bound states for energies $E<0$ and quasi-stationary states for
$0<E<V_{0}$.
The eigenvalues for the quasi-stationary states are obtained from the
condition that the wave function which is the solution of the
Schr\"odinger equation has the following asymptotic behavior~\cite{ref6}
\begin{equation}\label{eq2}
  \psi(r)\xrightarrow[r\rightarrow\infty]{}\frac{\text{e}^{ikr}}{r}.
\end{equation}
This produces discrete complex eigenvalues of the Hamiltonian. The
wave function that corresponds to this eigenvalue cannot be
normalized so it does not belong to the Hilbert space, but it has
several properties that permit us to interpret them as the wave
functions of the unstable state. The lifetime of the unstable
state is related to the imaginary part of the complex eigenvalue
$E_{0}-i\Gamma/2$ in the following way
\begin{equation}\label{eq3}
  \tau=\frac{\hbar}{\Gamma}.
\end{equation}

Another approach is to consider the scattering on the potential given
in Fig.~\ref{fig1}. Then one imposes the following asymptotic
condition that corresponds to the stationary state
\begin{equation}\label{eq4}
  \psi(r)\xrightarrow[r\rightarrow\infty]{}\sqrt{\frac{2}{\pi}}\sin(kr+\delta)
\end{equation}
and then from the condition that the wave function is regular inside
the well one obtains the $S$-matrix of the problem which has a simple
pole at a complex momentum at exactly the same value as in the first
approach. In this case the cross section for the scattering has the
Breit-Wigner form
\begin{equation}\label{eq5}
  \sigma\sim\frac{\Gamma^2/4}{(E-E_{0})^{2}+\Gamma^2/4}.
\end{equation}
When $\Gamma\neq0$ it is frequently stated that the
quasi-stationary state does not have the definite value of the
energy. It should however be stressed that in each single event
the energy is well determined. The meaning of the indefiniteness
of the energy of an unstable state is that in the range of the
energies $|E-E_{0}|\lesssim\Gamma$ the creation of the long lived
unstable state is possible. The other important point is the
scattering amplitude is given by the Breit-Wigner formula in the
vicinity of the pole which depends only on the position of the
pole and is not sensitive to other properties of the potential.
This property holds only for the energies close to the position of
the pole. For the energies $|E-E_{0}|>\Gamma$ the former statement
does not hold and one has to take into account other properties of
the potential. Also one has to say that there is no precise
division between the ranges of the energies where the Breit-Wigner
amplitude is sufficient for the description of the process and
where it is not, since there are always non-resonant contributions
in the process.
\section{Unstable states in quantum field theory}
In quantum mechanics the unstable state corresponds to a state of
the Hamiltonian that is ``almost'' bound. In other words the
unstable state has parts. In quantum field theory the unstable
states may be elementary. Let us consider the model presented by
Veltman~\cite{ref7} in which there are two scalar particles $A$
and $\phi$ whose masses fulfill the condition $M>2m$ ($M$ and $m$
are the masses of the $A$ and the $\phi$ fields, respectively).
The interaction Lagrangian of these fields is
\begin{equation}\label{eq6}
  {\cal L}_{I}=\frac{g}{2}[\phi^{2}(x)\cdot
  A(x)+A(x)\cdot\phi^{2}(x)].
\end{equation}
From the interaction Lagrangian it follows that the interactions
permit the transition
\begin{equation}\label{eq7}
  A\rightarrow\phi+\phi
\end{equation}
and from the condition on the masses $M>2m$ it follows that the decay
of the particle $A$ is kinematically possible so the particle $A$ is
unstable. Since $A$ is unstable it cannot appear as an asymptotic
state. Exclusion of some states from the asymptotic states can lead to
the
\begin{equation}\label{eq8}
  \text{\textbf{breakdown of the unitarity of the $S$ matrix}}.
\end{equation}
The danger is real. Veltman shows that there is no violation of the
unitarity of this theory if one includes only the states of the $\phi$
field as the asymptotic states. The idea of the proof is the
following.

In the lowest order of the perturbation theory the propagators of the
$\phi$ and $A$ fields are
\begin{align}\label{eq9}
  \Delta_{\phi}(x_{i}-x_{j})&=-\frac{1}{(2\pi)^{4}} \int d_4k\,
  \text{e}^{ik(x_{i}-x_{j})}\,\frac{1}{k^2-m^2+i\varepsilon},\\
\label{eq10}
\Delta_{A}(x_{i}-x_{j})&=-\frac{1}{(2\pi)^{4}} \int d_4k\,
\text{e}^{ik(x_{i}-x_{j})}\,\frac{1}{k^2-M^2+i\varepsilon}.
\end{align}
The propagator of $A$ in the next order is given by the diagram in
Fig.~\ref{fig2} and the analytic expression corresponding to this diagram has
the form
\begin{multline}\label{eq11}
  \Delta_{A}^{1}(x_{i}-x_{j})=\frac{1}{(2\pi)^{4}} \int d_4k\
  \text{e}^{ik(x_{i}-x_{j})}\,\{(k^{2}-M^{2})^{2}R_{2}(k^{2})\\
  +i\theta(k^{2}-4m^{2})I_{2}(k^{2})\}.
\end{multline}
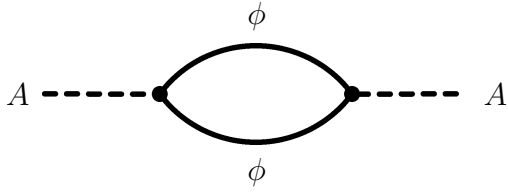
\begin{figure}[h]
\begin{center}
\begin{fmffile}{diagr1}
  \begin{fmfgraph*}(160,100) \fmfpen{thick} \fmfleft{i} \fmfright{o}
      \fmf{dashes}{i,v1} \fmf{dashes}{v2,o}
      \fmf{plain,left=0.5,tension=0.3,label=$\phi$}{v1,v2,v1}
      \fmflabel{$A$}{i} \fmflabel{$A$}{o} \fmfdot{v1,v2}
\end{fmfgraph*}
\end{fmffile}
\end{center}
\caption{Self energy diagram for the field $A$.}
\label{fig2}
\end{figure}
\noindent This diagram, after iteration, gives the geometric series with the
following factor
\begin{equation}\label{eq12}
  \frac{(k^{2}-M^{2})^{2}R_{2}(k^{2})
  +i\theta(k^{2}-4m^{2})I_{2}(k^{2})}{k^{2}-M^{2}+i\varepsilon}
\end{equation}
which has the singularity for $k^{2}\approx M^{2}$ and it appears
because of the condition $M>2m$. We thus see that the series is
divergent for any value of the coupling constant $g$ and the
perturbation method fails. The way out of this difficulty is to
find the propagator for such values of $k^{2}$ where the
perturbation series is convergent and then analytically continue
it to the vicinity of the point $k^{2}=M^{2}$. As the result one
obtains the following expressions for the propagators
\begin{equation}\label{eq13}
  \Delta_{\phi}(k^{2})=\frac{1}
  {k^{2}-m^{2}-(k^{2}-m^{2})^{2}R_{\phi}(k^{2})
    +i\theta(k^{2}-9m^{2})I_{\phi}(k^{2})},
\end{equation}
\begin{equation}\label{eq14}
  \Delta_{A}(k^{2})=\frac{1}
  {k^{2}-M^{2}-(k^{2}-M^{2})^{2}R_{A}(k^{2})
    +i\theta(k^{2}-4m^{2})I_{A}(k^{2})},
\end{equation}
where $m$ and $M$ are the physical masses. As can be seen the
propagator $\Delta_{\phi}(k^{2})$ has a pole for $k^{2}=m^{2}$ and
$\Delta_{A}(k^{2})$ \emph{does not} have a pole for $k^{2}=M^{2}$.
The fact that $\Delta_{A}(k^{2})$ is regular at the point
$k^{2}=M^{2}$ is crucial for the demonstration of the unitarity of the
$S$-matrix after the exclusion of the particles $A$ from the initial
and final asymptotic states. From these consideration we obtain the
following picture concerning the unstable states in quantum field
theory
\begin{itemize}
\item Unstable states should not be included as the initial and final
  asymptotic states.
\item $S$-matrix is unitary.
\item Unstable states appear only as the intermediate states and their
  propagators have the form dictated by the Dyson summation formula.
\end{itemize}
We thus see that the notion of the unstable state in quantum field
theory is realized through the elimination of the unstable states from
the asymptotic spaces and the modification of the propagators. In such
a way the space of the physical states is richer than the space of the
asymptotic states. The propagator of an unstable state does not have a
pole at the point $k^{2}=M^{2}$ but has a pole for a complex value of
$k^{2}$.

\begin{figure}[h]
  \label{fig3}
  \centering
  \begin{fmffile}{diagr2}
    \begin{fmfgraph*}(160,100) \fmfpen{thick} \fmfleft{i1,i2}
      \fmfright{o} \fmf{plain,label=$k_{1}$}{i1,v}
      \fmf{plain,label=$k_{2}$,label.side=right}{v,i2}
      \fmf{dashes,label=$k_{1}+k_{2}$}{v,o}\fmfdot{v}
      \fmflabel{$\phi$}{i1} \fmflabel{$\phi$}{i2} \fmflabel{$A$}{o}
    \end{fmfgraph*}
  \end{fmffile}
  \caption{Production of an unstable state $A$ by two particles
    $\phi$.}
\end{figure}
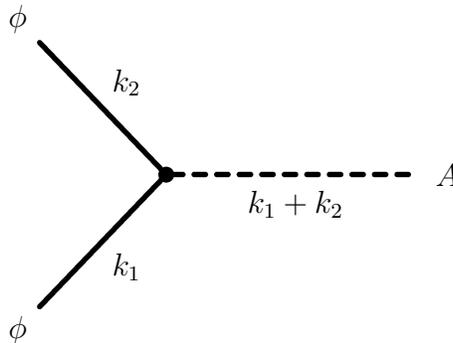

We thus see that in quantum field theory the vectors of the unstable
states are not introduced at all and unstable states are included as
the intermediate states with the special form of the propagators.
Despite this fact one can draw some conclusions about the properties
of the intermediate
states:\\[5pt]
\textbf{1. The momentum of the unstable states is well
  defined}\\[3pt]
From the general rules of quantum field theory the momentum is
conserved at each vertex (see e.g., Fig.~\ref{fig3}) so the momentum
of an unstable particle is well defined and is real.  \vspace*{10pt}
Later we will discuss the problem of the mass of an unstable state and
we will see that the mass of an unstable state can be for many
purposes considered as the complex quantity.\\[5pt]
\textbf{2. Gauge invariance}\\[3pt]
Gauge invariance is very important in elementary particle physics.
Quantum electrodynamics -- the most precise existing theory and
the standard model are examples of the theories with gauge
symmetry.  For this reason the gauge symmetry for the unstable
states has to be considered.

In the nineties the problem of the gauge invariance was widely
discussed in relation with the extremely precise measurement of
the $Z_{0}$ mass~\cite{ref9,ref8}. This discussion proliferated the
theoretical knowledge of unstable relativistic states and the
understanding of gauge invariance for such states. The accuracy of
the measurement of the $Z_{0}$ mass exceeded the accuracy of the
theoretical calculations of the next to the leading order in
perturbation theory and the next to the next to the leading order
calculations of mass and width were not gauge invariant. It is
clear that such a situation was very unsatisfactory both from the
theoretical and practical point of view, especially because the
gauge dependent corrections were not bounded and for some exotic
choice of the gauge they could be relatively large. The proposed
solution of this problem is to use for the mass definition the
$S$-matrix element of the corresponding process since this has
been proven to be gauge invariant~\cite{ref8a}. Therefore the pole
of the $S$-matrix should be used for the determination of the mass
and width of the resonances.

The problem of the gauge invariance and the correct form of the
propagator of an unstable state arose also for many processes
involving the gauge bosons $Z_{0}$ and $W^{\pm}$, where the
propagators appeared. The propagator of a stable vector particle
is given by
\begin{equation}\label{eq15}
  D_{\mu\nu}(q)=\frac{i\left(-g_{\mu\nu}+(1-\xi)
      \frac{q_{\mu}q_{\nu}}{q^{2}-\xi M^{2}}\right)}
  {q^{2}-M^{2}+i\varepsilon}
\end{equation}
where $\xi$ is the gauge parameter. This propagator has a pole for
$q^{2}=M^{2}$. For an unstable state the following substitution is
made in the denominator
\begin{equation}\label{eq16}
  q^{2}-M^{2}\rightarrow q^{2}-M^{2}+iM\Gamma
\end{equation}
and the following propagator is obtained
\begin{equation}\label{eq17}
  D^{\prime}_{\mu\nu}(q)=\frac{i\left(-g_{\mu\nu}+(1-\xi)
      \frac{q_{\mu}q_{\nu}}{q^{2}-\xi M^{2}}\right)}
  {q^{2}-M^{2}+iM\Gamma}
\end{equation}
where $\Gamma$ was called the width of the unstable state. It
turns out that such a propagator does not fulfill the Ward
identity so it is incompatible with the gauge invariance. The
correct form of the propagator must be the following
\begin{equation}\label{eq18}
  \tilde{D}_{\mu\nu}(q)=\frac{i\left(-g_{\mu\nu}+(1-\xi)
  \frac{q_{\mu}q_{\nu}}{q^{2}-\xi (M^{2}-iM\Gamma)}\right)}
  {q^{2}-M^{2}+iM\Gamma}
\end{equation}
Moreover it is not sufficient to insert the complex mass in the
propagator but it must also be inserted in all the vertices with
the $Z_{0}$ and $W^{\pm}$ bosons. This rule gives very important
information about the properties of the unstable states and is a
strong indication that the complex mass is the intrinsic property
of the unstable states and not only a mathematical trick. The
practical implementation of this idea is given in the rigged
Hilbert space quantum mechanics for the unstable states.
\section{Rigged Hilbert space of unstable states}
From the previous considerations we have seen that unstable states
appear both in non relativistic quantum mechanics and in the
relativistic quantum field theory. The description in these two cases
is drastically different.

In non relativistic quantum mechanics the unstable states are long
lived states described either as the eigenstates of the
Hamiltonian with complex eigenvalue of energy or as the stationary
eigenstates of the Hamiltonian for the scattering on the spherical
potential well. In both cases the eigenfunctions do not belong to
the Hilbert space so some of the axioms of the orthodox quantum
mechanics have to be relaxed.

In relativistic quantum field theory of~\cite{ref7} the situation is
completely different. The unstable particle or the resonance is
elementary and its field appears in the Lagrangian of the system. The
kinematical mass relation allows the decay of the unstable elementary
state. The elementary unstable particles are \emph{eliminated} from
the set of asymptotic states. Such a theory remains unitary provided
the modification of the propagator of the unstable state is done
according to the Dyson expansion. It should be noted that in such an
approach the notion of a vector for an unstable state does not exist
in quantum field theory: the unstable states appear only as
intermediate states and one has to use the right form of the
propagator for them. This is the only way how the unstable states
appear in the theory. The important feature of this approach is that
the kinematical mass of an unstable state is complex and such a mass
should be used without exception not only in the construction of the
propagators but also in the vertices. The complex mass is taken from
the position of the pole of the $S$-matrix.

It is the natural now to ask the question: does there exist a
unified formalism that is mathematically precise and has the
physical properties that were required by the two approaches
discussed earlier? The formalism with such properties is
implemented by the rigged Hilbert space quantum mechanics whose
principal properties are the following~\cite{ref10}
\begin{itemize}
\item The linear space for states and observables is provided by
the rigged Hilbert space which is a triplet of the spaces, one of
which is the ordinary Hilbert space of the system, the two others
is a (dense) subspace of it and the dual thereof containing the
(Dirac) kets and other generalized vectors like Gamow states.
\item The dynamical differential equations of motion are identical
to those in conventional quantum mechanics but their boundary
conditions are not the Hilbert space conditions but given by the
dense subspace.
\item The conventional quantum mechanics is
contained as a ``limiting'' case within the rigged Hilbert space
quantum mechanics.
\end{itemize}
The rigged Hilbert space quantum mechanics which is an extension of
the conventional quantum mechanics describes the wider class of the
physical phenomena and also adds mathematical precision to the
conventional quantum mechanics. The main important results of the
rigged Hilbert space quantum mechanics are~\cite{ref10,ref11,ref12}
\begin{itemize}
\item Dirac formalism of bras and kets.
\item Precise meaning of the Lippmann-Schwinger kets.
\item Description of the unstable states.
\end{itemize}

I will briefly discuss here only the problem of the unstable states.
The unstable states appear only in scattering as intermediate states.
In terms of the energy wave function the rigged Hilbert space consists
of the following three spaces: the Hilbert space $\mathcal{H}$ is the
space $L^{2}$ of the Lebesgue square integrable functions.  For the
definition of the rigged Hilbert space one has to choose a linear
space $\Phi$ with stronger than Hilbert space convergence (topology)
which is dense in the Hilbert space. Then together with the dual space
they form the rigged Hilbert space
\begin{equation}\label{eq19}
  \Phi\subset{\cal H}\subset\Phi^{\times}.
\end{equation}
Usually, for Dirac kets, one chooses for $\Phi$ the Schwartz space.
For unstable states the choice of the RHS is $\Phi_{+}\subset{\cal
  H}\subset\Phi_{+}^{\times}$ where the space $\Phi_{+}$ is in the
energy representation the space ${\cal H}^{2}_{+}\cap{\cal
  S}|_{{\mathbb R}_{+}}$ where ${\cal H}^{2}_{+}$ is the space of the
Hardy class functions in the upper complex half plane and ${\cal S}$
is the Schwartz function space.

The state of the unstable particle is defined in terms of the
Lippmann-Schwinger kets $|[j,\textsf{s}_{\text{R}}]b^{-}\rangle \in
\Phi_{+}^{\times}$
\begin{equation}\label{eq20}
  |[j,\textsf{s}_{\text{R}}]b^{-}\rangle=\frac{i}{2\pi}\int_{-\infty}^{\infty}
  |[j,\textsf{s}]b^{-}\rangle\frac{1}{\textsf{s}-\textsf{s}_{\text{R}}}.
\end{equation}
Here $|[j,\textsf{s}_{\text{R}}]b^{-}\rangle$ denotes the state vector
of the unstable particle (Gamow ket) and $|[j,\textsf{s}]b^{-}\rangle$
is the Lippmann-Schwinger ket, $j$ is the spin, $\textsf{s}$ the
energy square and $b$ are other quantum numbers. The complex value
$\textsf{s}_{\text{R}}$ is the position of the $S$-matrix pole on the
second sheet of the lower complex energy plane. The state~\eqref{eq20}
may be considered as the generalization of the non-relativistic wave
function~\cite{ref1} and its definition here is fully relativistic.
The vector defined by~\eqref{eq20} is a (generalized) eigenvector of
the selfadjoint operator $M^{2}=P^{\mu}P_{\mu}$ with
$\textsf{s}_{\text{R}}$ as its eigenvalue
\begin{equation}\label{eq21}
  P^{\mu}P_{\mu}|[j,\textsf{s}_{\text{R}}]b^{-}\rangle=
  \textsf{s}_{\text{R}}|[j,\textsf{s}_{\text{R}}]b^{-}\rangle
\end{equation}
which means that the square of the mass of the unstable state is a
complex number -- the position of the pole of the $S$-matrix. The
situation here is identical as in the case of quantum field
theory. The property~\eqref{eq21} is a justification of the
quantum field theory rule~\eqref{eq16}.

The important property of the Gamow vectors~\eqref{eq20} is its
time evolution. Here it turns out that the time evolution of the
vector~\eqref{eq20} is permitted only for $t\geq0$~\cite{ref11}
and is given by
\begin{equation}\label{eq22}
  \text{e}^{-iH^{\times}t}|[j,\textsf{s}_{\text{R}}]b^{-}\rangle =
  \text{e}^{-iE_{\text{R}}t}\text{e}^{-\Gamma_{\text{R}}t/2}
  |[j,\textsf{s}_{\text{R}}]b^{-}\rangle\qquad\text{for
  $t\geq0$}.
\end{equation}
where $E_{\text{R}}$ and $\Gamma_{\text{R}}$ are the following
parameterizations of the complex $\textsf{s}_{\text{R}}$ in terms
of two real numbers
\begin{equation}
\label{eq22a}
 \textsf{s}_{\text{R}}=\left(
E_{\text{R}}-i\frac{\Gamma_{\text{R}}}{2}\right)^{2}.
\end{equation}
This is a different parameterization of the complex mass than the one
used in~\eqref{eq16}--\eqref{eq18}.  From~\eqref{eq22} follows that
the decay probability of the Gamow ket is exponential and the lifetime
of this exponential decay is $\tau=\frac{1}{\Gamma_{\text{R}}}$.
Therefore~\eqref{eq22a} is a better parameterization
than~\eqref{eq16}.  The vectors~\eqref{eq20} form the semigroup
representation of the Poincar\'e group transformations into the
forward light cone, they are classified by
$[j,\textsf{s}_{\text{R}}]$~\cite{ref13}. This representation is the
minimally complex representation where the momentum is complex but the
velocity is real. The transformation property under the Poincar\'e
semigroup transformation, of which Eq.~\eqref{eq22} is the special
case, introduces a new quantum mechanical arrow of time which
distinguishes the semigroup time evolution of scattering and decay
theory based on~~\eqref{eq19} from the unitary Poincar\'e group
evolution for the asymptotic states.

\section{Conclusions}
I have presented here the original quantum mechanical and field
theoretic approaches to the problem of unstable states and compared
them to the rigged Hilbert method. My discussion can be summarized as
follows
\begin{enumerate}
\item The vectors of the unstable states used in the conventional
  textbook treatments of this subject do not belong to the Hilbert
  space.
\item In quantum field theory there are no vectors corresponding to
  the unstable states. Resonances are included in the theory only as
  intermediate states with the special form of the propagator that
  corresponds to the complex mass.
\item The rigged Hilbert space quantum mechanics gives the precise
  mathematical meaning to the vectors of the unstable states in
  quantum mechanics.
\item The RHS vectors for the unstable states have a complex mass
  eigenvalue.  This explains the field theoretic rule
  \[
  M^{2}\rightarrow M^{2}-iM\Gamma
  =\left(M_{R}-i\frac{\Gamma_{R}}{2}\right)^{2}.
  \]
\item The rigged Hilbert space quantum mechanics predicts new
  phenomena like the new quantum mechanical arrow of time.
\end{enumerate}
\section*{Acknowledgment}
The author thanks Dr.\ L\'{\i}dia Ferreira, the organizer of the CFIF
Workshop ``Time Asymmetric Quantum Theory: the Theory of Resonances'',
Lisbon, Portugal, 2003 for her warm hospitality and creation of an
excellent working environment during the conference.

\typeout{  }
\typeout{  }
\typeout{  }
\typeout{!!!!!!!!!!!!!!}
\typeout{first run latex twice on the main file, then}
\typeout{run metafont with the following commands}
\typeout{mf '\protect\mode:=cx; input diagr1'}
\typeout{mf '\protect\mode:=cx; input diagr2'}
\typeout{(the metafont mode cx depends on the local printer!!)}
\typeout{and then run latex again on the main file}
\typeout{!!!!!!!!!!!!!!}
\typeout{  }
\typeout{  }
\typeout{  }
\end{document}